\documentclass[10pt,journal,letterpaper]{IEEEtran}
\usepackage[utf8]{inputenc}
\usepackage{multirow}
\usepackage{makecell}
\usepackage{array}
\usepackage{graphicx}
\usepackage{float}
\usepackage{amssymb}
\usepackage{color}

\begin{document}

\title{Enabling Cognitive Smart Cities Using  Big Data and Machine Learning: Approaches and Challenges}

\author{Mehdi Mohammadi,~\IEEEmembership{Graduate Student Member,~IEEE,} Ala Al-Fuqaha,~\IEEEmembership{Senior Member,~IEEE}
\IEEEcompsocitemizethanks{\IEEEcompsocthanksitem Mehdi Mohammadi and Ala Al-Fuqaha are with the Department of Computer Science, Western Michigan University, Kalamazoo, MI 49008 USA.
\protect
E-mail: \{mehdi.mohammadi,ala.al-fuqaha\}@wmich.edu.
}

}

\markboth{IEEE Communications Magazine,~Vol.~x, No.~x, xxxxx~201x}
{Shell \MakeLowercase{\textit{et al.}}: Bare Demo of IEEEtran.cls for Computer Society Journals}

\IEEEcompsoctitleabstractindextext{
\begin{abstract}
The development of smart cities and their fast-paced deployment is resulting in the generation of large quantities of data at unprecedented rates. Unfortunately, most of the generated data is wasted without extracting potentially useful information and knowledge because of the lack of established mechanisms and standards that benefit from the availability of such data. Moreover, the high dynamical nature of smart cities calls for new generation of machine learning approaches that are flexible and adaptable to cope with the dynamicity of data to perform analytics and learn from real-time data.
In this article, we shed the light on the challenge of under utilizing the big data generated by smart cities from a machine learning perspective. Especially, we present the phenomenon of wasting unlabeled data. We argue that semi-supervision is a \emph{must} for smart city to address this challenge. 
 We also propose a three-level learning framework for smart cities that matches the hierarchical nature of big data generated by smart cities with a goal of providing different levels of knowledge abstractions.
The proposed framework is scalable to meet the needs of smart city services. Fundamentally, the framework benefits from semi-supervised deep reinforcement learning where a small amount of data that has users' feedback serves as labeled data while a larger amount is without such users' feedback serves as unlabeled data. The framework utilizes a mix of labeled and unlabeled data to converge toward better control policies instead of wasting the unlabeled data.
This paper also explores how deep reinforcement learning and its shift toward semi-supervision can handle the cognitive side of smart city services and improve their performance by providing several use cases spanning the different domains of smart cities. We also
highlight several challenges as well as promising future research directions for incorporating machine learning and high-level intelligence into smart city services.
\end{abstract}
}

\maketitle
\IEEEdisplaynotcompsoctitleabstractindextext
\IEEEpeerreviewmaketitle

\section{Introduction}\label{sec:Introduction}

\begin{figure}
	\begin{center}
		\includegraphics[width=0.5\textwidth]{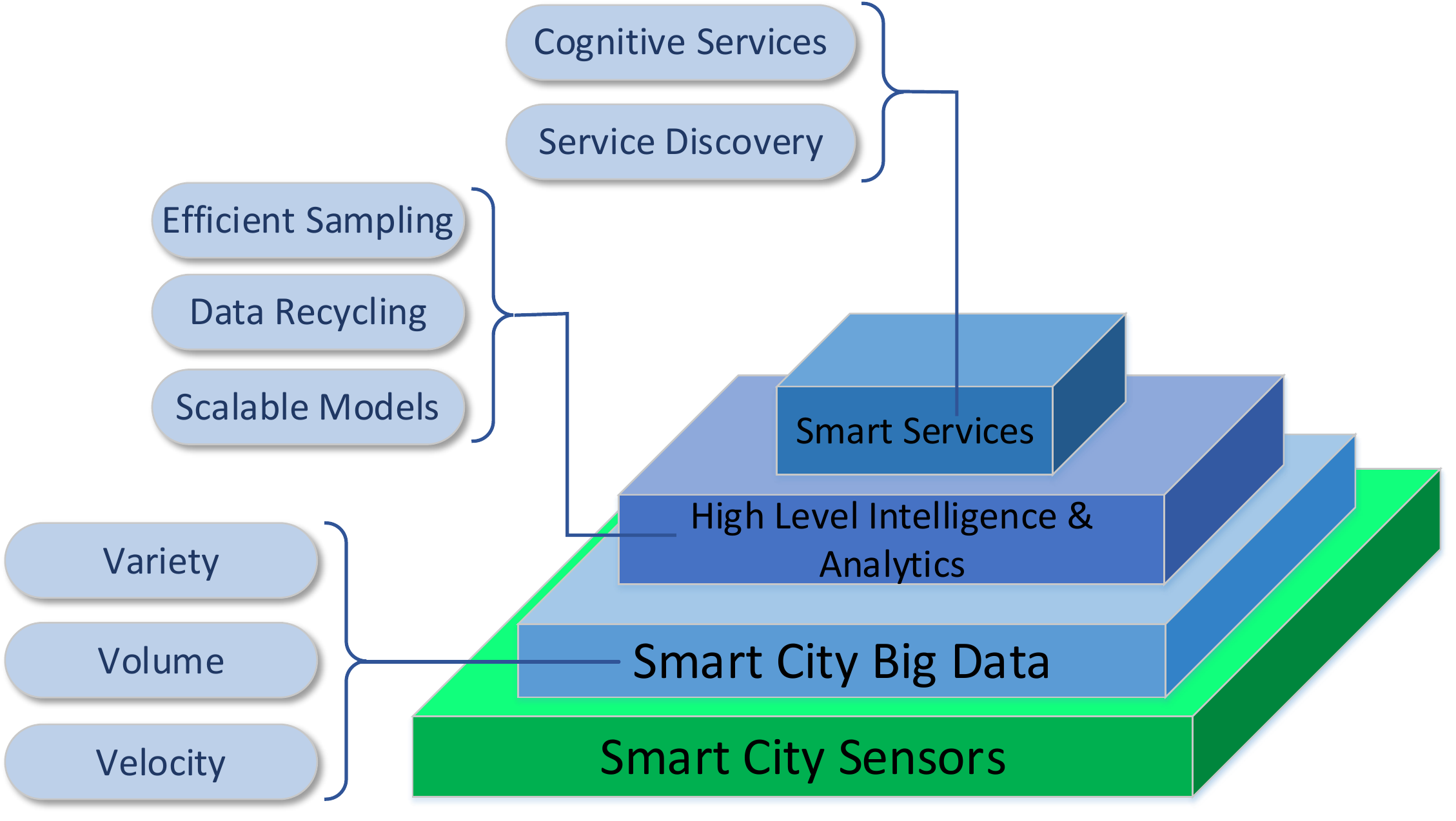}

	\end{center}
	\caption{Challenges of Smart Cities from a Machine Learning perspective.} \label{fig:mainIdea}
\end{figure}

Smart cities provide services that benefit from the city-scale deployment of sensors, actuators and smart objects \cite{ala2015internet}. 
Such services are mainly driven by data and can be broadly classified as producers of data, consumers of data, or a combination of both. For example, a parking service that deploys Message Queue Telemetry Transport (MQTT) broker to publish the parking lots' availability data is considered a producer while cars that subscribe to that broker are considered as consumers. Cars can be producing other data for use by other smart city components. For instance, cars use Device-to-Device (D2D) communications to alert nearby vehicles and pedestrians of their presence and potential traffic hazards. In a city-scale deployment of smart services, data is generated at high rates which presents new challenges for smart city designers and developers. Beyond the challenges for data management of big data represented by 3V's (volume, variety, and velocity), there are other challenges from  analytics and Machine Learning (ML) perspectives (cf. Figure \ref{fig:mainIdea}). Unfortunately, only a small fraction of the massive smart city data is typically utilized by smart services to improve the lives of the city's residents. The main culprit is the lack of a large amount of labeled data. This calls for the need to utilize machine learning algorithms that exploit the availability of unlabeled and labeled data in the context of smart cities. 

Analogous to  the waste recycling processes and standards in urban cities, there is a need for processes and mechanisms for data recycling in smart cities where hundreds or thousands of Gigabytes of data is generated per second. Data analytic methods and machine learning algorithms should be able to extract knowledge and useful information from data to reduce the amount of \emph{digital waste}.

Despite the recent advancements in computing and storage technologies, most of the data analytic approaches exploit sampling methods that are efficient in terms of time complexity but neglect a large part of data that may contain important patterns that are not represented by the samples. On the other hand, through the use of Deep Neural Networks (DNNs), datasets with millions of parameters can be considered to extract insightful analytics. 

Anecdotal data indicates that when smart city data is not used for learning and analytics in a short-term, it is unlikely that it would be used later. It is estimated that by 2012 only about 0.5\% of all 2.8 Zettabytes (ZB) of stored data have been analyzed and 3\% of them are labeled based on a study by IDC\footnote{https://www.emc.com/collateral/analyst-reports/idc-the-digital-universe-in-2020.pdf}. This highlights the challenge of potentially wasting hidden information in 99.5\% of the generated data.

Lately, there have been active discussions on the governance, management, and storage of smart cities big data, but there is no clear answer on how to use the enormous amounts of collected data. Should it be directly incorporated into analytics and machine learning activities? Or should it be sampled? Even though in many cases sampling approximates the solution, for smart city services where the preference of citizens comes to play, or individual activities affect the whole community, sampling may not be ideal. For example, to predict anomalies in a city's infrastructure considering all collected data from the various sensor sources would help to realize such services. Another example includes services that predict criminal or discriminatory activities through social media comments (e.g., Tweeter, Facebook, etc.). For such services, considering a wider range of data is necessary, since the criminal or discriminatory comments may constitute a small portion of the whole data.

Smart city ecosystems have the following characteristics from a machine learning perspective:

\begin{itemize}
\item Humans need to interact with the systems to provide their feedback. 
\item Many sensors and smart devices generate data at a high rate. Not all the data can be reviewed by humans for justification, but the system should learn and improve itself from previous experiences. 
\item They need a general, dynamic, and continuous learning mechanism as the context of a smart city application is not always fixed and the operating environment of smart city applications evolves over the time.
\item The data generated by smart city applications is noisy or has some degree of uncertainty.
\end{itemize}

Based on these characteristics, we believe an integration of DNNs, Reinforcement Learning (RL), and semi-supervised learning can address these issues and deliver complete adaptive solutions. 

The need for deep learning approaches stems from the  need to extract high-level abstractions from the raw data. Each layer of a DNN generates an abstract representation of its input data. To get more levels of abstraction, more hidden layers of neurons are needed.

Reinforcement learning has been studied well for control systems and systems that need to perform autonomic actions. In reinforcement learning, there is no output (i.e., classification) for the training data - which is the case for many smart city applications- instead, choosing the right actions is rewarded. The goal of a RL system is to find an action for each state of the system such that the total reward of the learning agent is maximized in the long-term. On the other hand, it is infeasible or extremely tedious for the users to provide a reward feedback for all the training data. This issue can be addressed through the use of semi-supervised learning approaches where data is partially labeled.

Semi-supervised machine learning approaches are a promising method to address the scarcity of annotated data in big data streams. Moreover, Deep Reinforcement Learning (DRL) approaches also have shown promising results in systems where a reward feedback from the environment is needed to improve the performance of the system instead of a class label as in the case of supervised learning methods. The combination of these techniques can help to extract the most value from the big data generated by smart cities.

In our proposed method, we combine the strength of these approaches for delivering semi-supervised DRL agents that learn from the smart cities data to perform the best actions on the environment. The proposed approach is an enabler for cognitive smart city services since the learning agent evolves as the conditions of the environment change, and performs autonomic actions without human interventions.

\begin{figure*}
	\begin{center}
		\includegraphics[width=1\textwidth]{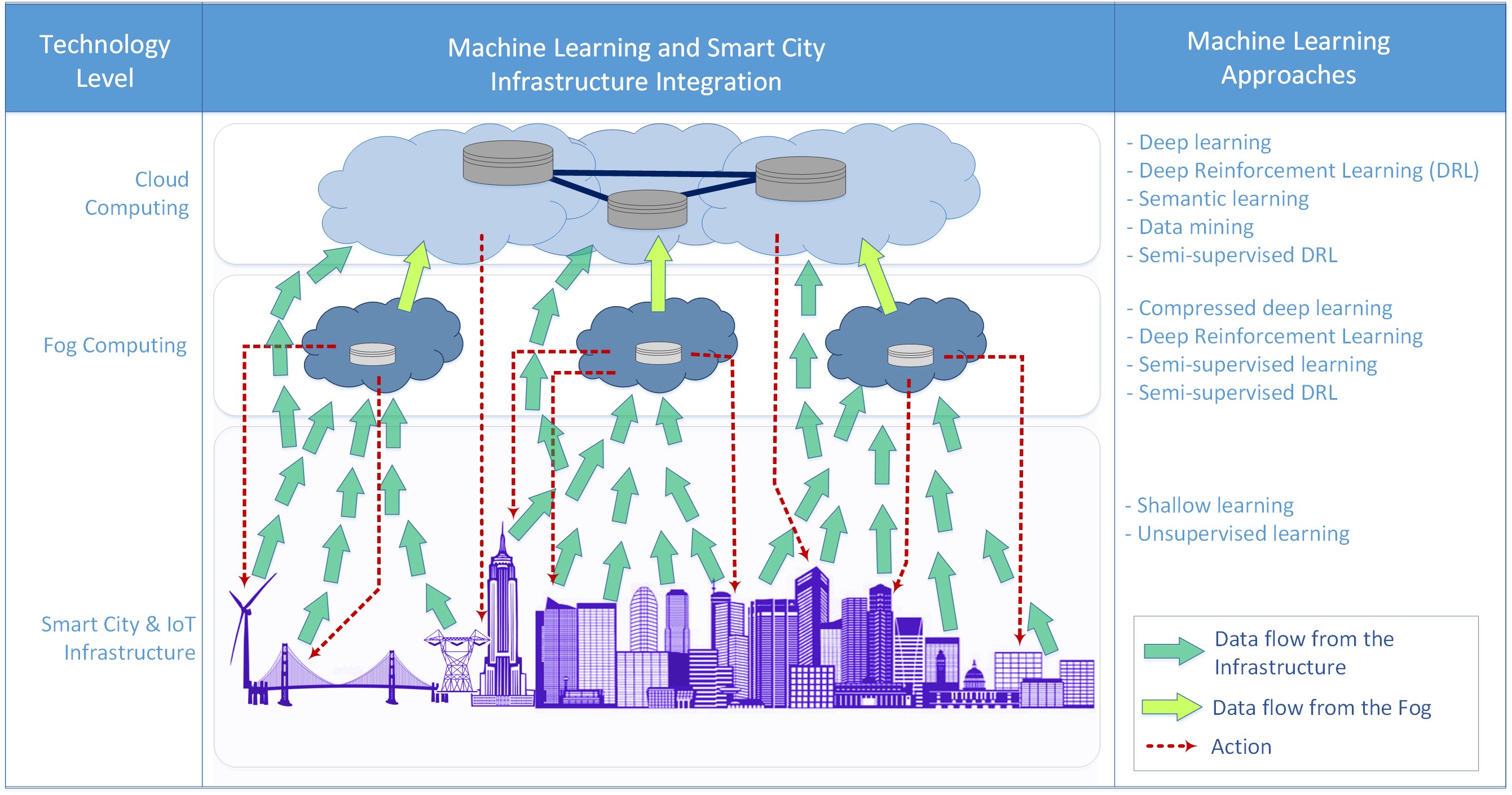}

	\end{center}
	\caption{The levels of intelligence in smart cities.} \label{fig:iotInt}
\end{figure*}

\section{Related Work}\label{sec:relatedWork}

\emph{Cognitive smart city} refers to the convergence of emerging IoT and smart city technologies, their generated big data, and artificial intelligence techniques. 
Among the commercial products that move toward cognitive frameworks, IBM Watson offers a cognitive system with several analytic and machine learning services that rely on dynamic learning (i.e., the learning process is improved in future rounds based on the feedback from previous rounds).
Cognitive computing is a term used by IBM to describe systems that can learn from a wide range of datasets, are able to provide reasons, can interact with humans through natural languages, and gain their experiences in the context.

Google Now is another service for making suggestions to the user, and bringing the most useful information to the user at the right time and place. This system learns from the users' past behavior and input in their Google accounts such as Calendar, Chrome, Gmail, Search, and Youtube \cite{kaltenrieder2015fuzzy}. Its use of natural language understanding integrated with other services such as search engine poses it closer to the cognitive era.

Haven OnDemand\footnote{http://www.havenondemand.com} is a machine learning-centric development platform by HP Enterprise that provides APIs for creating cognitive services. Text analysis, speech recognition, image analysis, indexing and search are among the APIs that are offered to developers.

In the research community, there are several works that propose cognitive solutions that fit the needs of IoT-based systems.
Vlacheas \textit{et al.} \cite{vlacheas2013enabling} proposed a cognitive management framework in the context of smart cities to enable smart objects to connect to the most relevant objects and consequently bring more value to the end user. In their framework, they focused on the reuse of the functionality and services of available objects through three levels including virtual objects (VO), composite virtual objects (CVO), and service level. The service level derives the functionality of the requested service that is required by a stakeholder or a given application. These functionalities are delegated to CVOs to be carried out. The authors showed that the service execution time in their proposed framework is decreased leading to lower operational expenditures.

Cognitive Internet of Things (CIoT) \cite{wu2014cognitive} is another research that was conducted by Wu \textit{et al.} to deliver a cognitive framework for IoT applications. The framework offers interactions between five cognitive tasks including: perception-action cycle, massive data analytics, semantic derivation and knowledge discovery, intelligent decision-making, and on-demand service provisioning. They identified two areas that are required for objects in a cognitive environment to understand and learn, namely: derive the semantic from analyzed data, and discover valuable patterns and rules as knowledge.

The authors in \cite{feng2017smart} defined a cognitive framework for smart homes based on cognitive dynamic systems and IoT. In the core of their cognitive memory, they used a Bayesian model, a Bayesian filter, and reinforcement learning. The Bayesian model is placed on top of perceptor which observes the environment. The Bayesian filter estimates the state of the system and reinforcement learning provides the mechanism to choose the best possible actions based on the total received rewards.

In contrast to centralized intelligence and analytics on the cloud, the authors in \cite{tang2017incorporating} proposed to integrate artificial intelligence in fog computing to facilitate smart city big data analysis. They introduced a hierarchical fog computing model to analyze big data for smart city applications. Using this model, the overall performance is enhanced through reducing the communications bandwidth by not having to transmit all raw data to the cloud, and performing real-time analytics due to the closeness of the fog to the source of data. They used Hidden Markov Model (HMM) approach in their model to support big data analysis in a smart pipeline monitoring system.

Table \ref{tbl:relatedWork} summarizes the works in this field and shows which levels of big data generation are covered by their intelligence and analytics. It also indicates the position of this study relative to these works. Compared to the aforementioned works that bring analytics to the fog or cloud levels, our approach aims to deploy analytic solutions on the fog and the cloud, which in turn covers a large number of smart city applications including time-sensitive and non-time-sensitive ones. Moreover, in order to improve the accuracy of the analytics, the proposed approach digs into the larger body of data where data is untapped and no labels or meta-data are provided.

\begin{table*}[]
\renewcommand{\arraystretch}{1.35}
\centering
\caption{Support of Machine Learning Intelligence in Smart city Context.}
\label{tbl:relatedWork}
\begin{tabular}{|l|l|c|c|c|c|c|c|}
\hline
\multicolumn{2}{|l|}{\multirow{2}{*}{Related Work}}                                                   & \multicolumn{3}{c|}{ML Support level}    & \multirow{2}{*}{Domain} & \multirow{2}{*}{ML Algorithm} & \multirow{2}{*}{Use Case} \\ \cline{3-5}
\multicolumn{2}{|l|}{}                                                                                & Infrastructure & Fog        & Cloud      &                         &                               &                           \\ \Xhline{3\arrayrulewidth}

\multirow{3}{*}{Commercial} & \textbf{IBM Watson}   &             &                  & \checkmark & General      & Various            &    Healthcare, Crime detection              \\ \cline{2-8} 
                            & \textbf{Google Now}   &             &                  & \checkmark  & User-centric      & Various        &   Traffic,  Transit                  \\ \cline{2-8} 
                            & \textbf{HPE Haven OnDemand}             &              &                   & \checkmark  & General      &  Various     &   Sentiment Analysis                     \\ \hline
\multirow{4}{*}{Research}   & \textbf{\makecell[l]{Cognitive management\\ framework \cite{vlacheas2013enabling}}}            & \checkmark             & \checkmark                  &                             & Smart city      & \makecell{Pattern Recognition,\\ Semantic Reasoning}  & \makecell{Smart health,\\ Public
safety,\\ Smart transportation} \\ \cline{2-8} 
                            & \textbf{Cognitive IoT \cite{wu2014cognitive}}         &              &                   & \checkmark                           & Smart city      & Multiagent Learning  &  \makecell{Convenient smart home,\\ Real-time traffic routing} \\ \cline{2-8}
							& \textbf{\makecell[l]{Cognitive interactive\\ framework\cite{feng2017smart}}}         &              & \checkmark                  &                            & Smart home      & Reinforcement Learning  & Convenient smart home \\ \cline{2-8}
                            & \textbf{Intelligence in Fog \cite{tang2017incorporating}}            &              & \checkmark                  &                         & Smart city      & HMM   & Smart pipeline   \\ \cline{2-8}
                            & \textbf{Intelligent gateway \cite{ala2015toward}}            &              & \checkmark                  &                              & IoT      & Rule-based  & Smart healthcare \\ \cline{2-8}
                            & \textbf{Current work} &              & \checkmark                 & \checkmark                            & Smart city      & \makecell{Semi-supervised Deep\\ Reinforcement Learning}  & \makecell{Energy, Water,\\ Agriculture,\\ Transportation,\\ Healthcare}  \\ \hline
\end{tabular}
\end{table*}

\section{Intelligence for Smart Cities}\label{sec:intelligence4smartcity}

In this section, we introduce the overall framework for intelligence in smart cities. The framework offers three levels of intelligence, namely: the level of \emph{smart city and IoT infrastructure}, \emph{fog computing}, and \emph{cloud computing}. Figure \ref{fig:iotInt} illustrates the overall position of machine learning approaches within the hierarchy of smart city infrastructure where each component of the smart city system is controlled by an intelligent software agent which is deployed in the fog or the cloud depending on the characteristics of the required analytics (e.g., time-sensitiveness). Consequently, raw data can be transferred to the fog or to the cloud. The running analytics agent then returns an appropriate action to the infrastructure devices based on predictions (e.g., adjust traffic light timing based on traffic congestion data from the corresponding roads).

The motivation behind this architecture is that deeper levels of data abstraction and knowledge representation can be obtained as the data travels through the smart city infrastructure. At the highest level, a city-wide abstraction is needed to manage the city's resources and services on a long-term basis. On the other hand, at the lowest level,  sensors and smart objects generated data is used to manage the resources and services on a short-term basis. Moreover, fog-based analytics support local actions in predefined contexts, while cloud-based analytics are capable of covering larger geographical regions with various contexts.

The level of IoT infrastructure is where the sensors and resource-constrained devices percept the environment. The resource limitation of these devices inhibits the deployment of complex and large learning models. Instead, several shallow machine learning approaches, including unsupervised and semi-supervised methods (e.g., K-Nearest Neighbors, Support Vector Machines, etc.) can be applied in the context of these devices to make them smart. However, to bring analytics and intelligence closer to the source of data (e.g., end users, IoT resource-constrained devices) there is a need to utilize modern and advanced learning models like deep learning. A nascent research path is to overcome the resource limitation of these devices to allow them to utilize deeper neural network models. In recent years, several approaches have been proposed to compress or prune deep neural networks so that they can be loaded into IoT resource-constrained devices, wearable electronics, and smart phones~\cite{han2015learning}. Using such compressed neural networks, it is possible to integrate deep reinforcement learning with these devices.   

At the fog computing level, the raw data is aggregated and transmitted to the cloud computing level. Compressed deep learning models, DRL, and semi-supervised methods can be used at this level as the resources at this level have lesser constraints compared to the IoT resources. The proposed semi-supervised DRL approach is also applicable at this level. Also at this level, light-weight intelligence needs to be brought to the IoT gateways and proxies to enable the efficient realization of horizontal integration of services in support of smart city applications \cite{ala2015toward}.

At the cloud computing level, more complex and large-scale machine learning and data mining frameworks and algorithms can be integrated with semantic learning and ontologies to extract high-level insights and patterns from the collected data. Deep learning models are highly fit at this level as they are able to provide deeper abstractions of the data. Recent advancements in Graphics Processing Unit (GPU) technology as well as the development of efficient neural network parameter initialization algorithms (e.g., autoencoders), the use of Rectified Linear Units (ReLUs), and the introduction of Long Short-Term Memory (LSTM) neural networks and their variants, help to solve the vanishing gradient problem; therefore, allowing the realization of efficient deeper learning models.  

\section{Emerging Approaches}\label{sec:approaches}

Reinforcement learning aims to imitate the learning process of humans. Through the reinforcement learning method, an agent can sense the environment through several sensor inputs. The agent uses these raw inputs to generalize the experience of the system for confronting new and unknown situations. The combination of reinforcement learning and deep neural networks - known as Deep Reinforcement Learning - has resolved several limitations of reinforcement learning including the limitation in the diversity of application domains, the need for manual engineering features, and their poor scalability for high-dimensional state-space domains \cite{mnih2015human}.

\begin{figure}
	\begin{center}
		\includegraphics[width=0.45\textwidth]{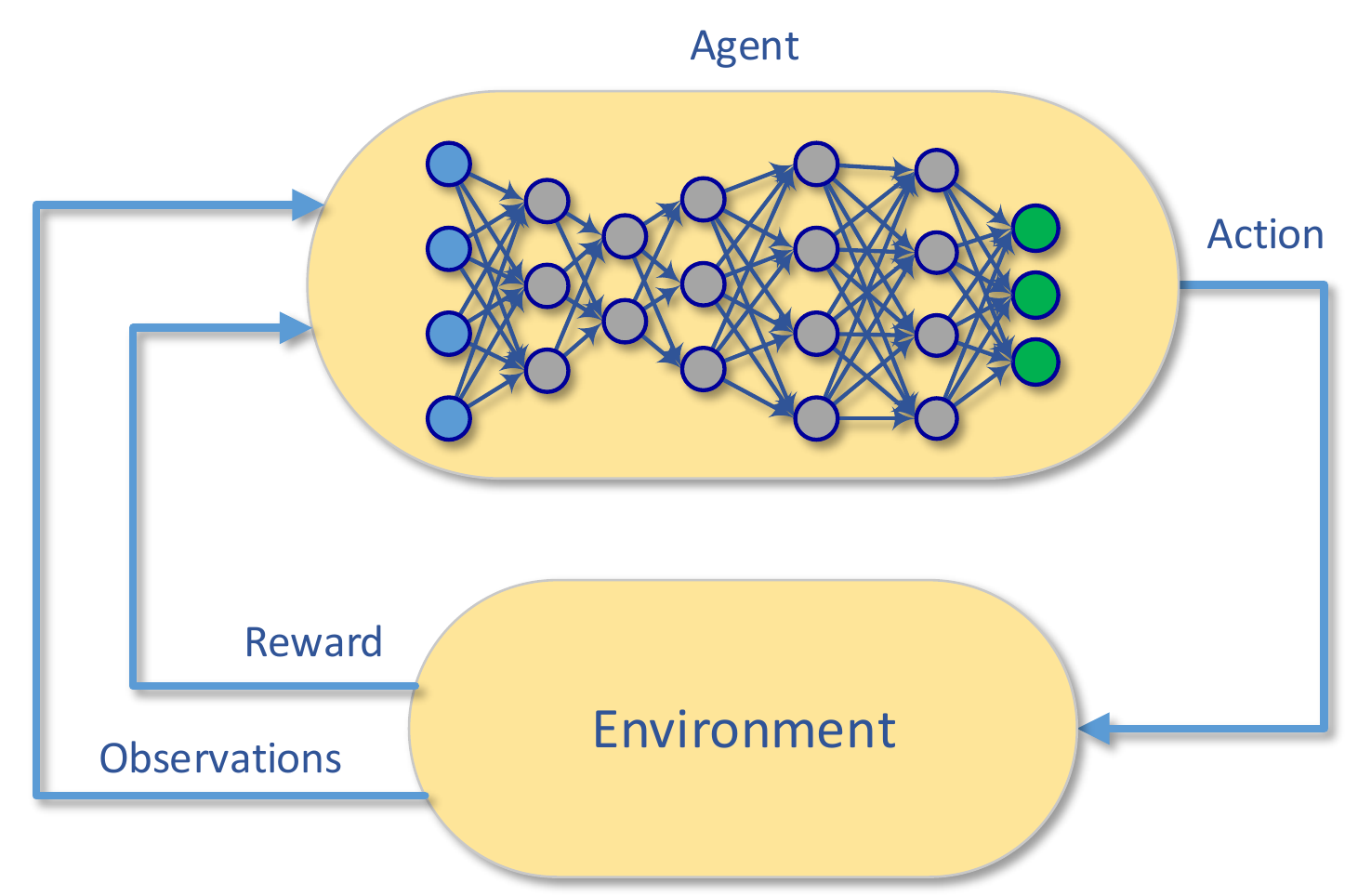}

	\end{center}
	\caption{A conceptual structure of a deep reinforcement learning system.} \label{fig:DRL_Model}
\end{figure}

A DRL agent observes the environment parameters, takes actions on the environment, and receives a reward feedback for each action. The objective of the agent is to maximize its total future rewards. A deep neural network is used to approximate the optimal action-value function (i.e., which action is best to pick for a given state to maximize future rewards). Figure \ref{fig:DRL_Model} illustrates the high-level conceptual structure of a DRL system.

In our proposed semi-supervised DRL approach, we adapted generative deep neural networks (e.g., Variational Autoencoder-VAE) \cite{kingma2014semi} as the semi-supervised component of the model. In our proposed model, VAE is extended to produce the probability of each action in the system.

The proposed approach was applied in a smart campus project as part of a smart city \cite{mohammadi2017semi}. The objective of the semi-supervised learning agent is to provide indoor localization and navigation services where its reward function is defined to be the reciprocal of the distance to the target point. The agent learns from the fingerprints of RSSI readings of several Bluetooth Low Energy (BLE) iBeacons in the environment to take best actions (i.e., moving north, west, etc.). We had a dataset of RSSI fingerprints in which 15\% of data points were annotated with the location details (i.e., labeled data).  

\begin{figure}
	\begin{center}		
    
    		\includegraphics[width=0.5\textwidth]{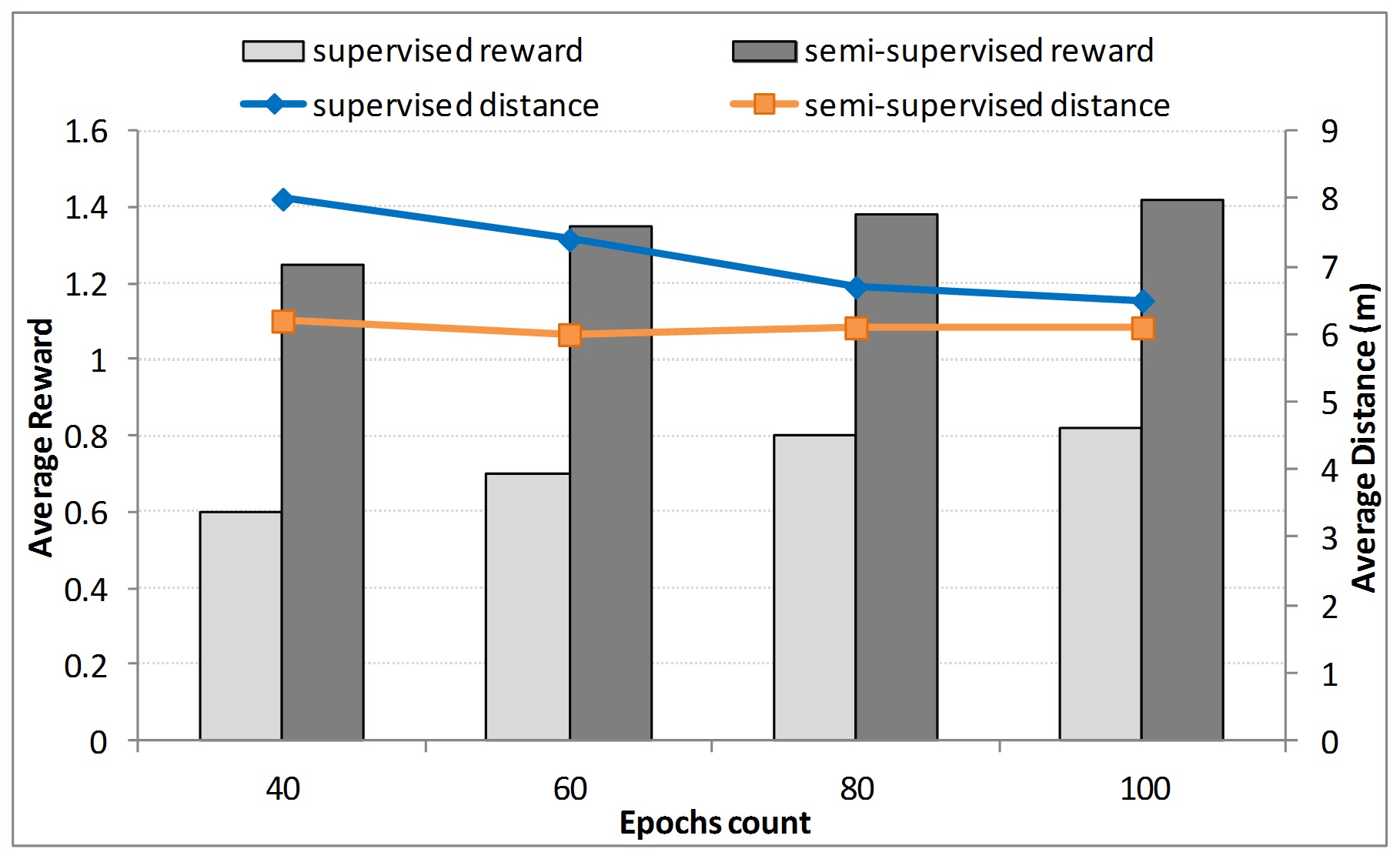}
	\end{center}
	
	\caption{The performance of semi-supervised DRL model versus supervised DRL model showing the average total rewards and the average distance to the target over different epoch counts.}
    \label{fig:reward_comp}
	
\end{figure}

Figure \ref{fig:reward_comp} compares the supervised and semi-supervised DRL models. The results indicate that using the semi-supervised DRL model that uses the combination of labeled and unlabeled data improves the performance of the system.  From the total rewards point of view, the semi-supervised model achieves higher rewards quickly compared to the supervised model. It gains between 60\% to 100\% more rewards compared to the supervised model gains in different number of epochs. In terms of accuracy of localization, the semi-supervised model reaches closer to the target point achieving an improvement between 6\% to 23\% compared to the accuracy of the supervised model.

The proposed semi-supervised DRL model can serve the applications in the fog and cloud layers of a smart city since the underlying deep neural network would be complex and large depending on the type of application and cannot reside on IoT resource-constrained devices. However, more investigation is needed to bring this algorithm to IoT devices.

\section{Smart City Use Cases}\label{sec:FEW_nexus}

When we think about a smart city where the management and control of the city's resources is performed through intelligent information systems, we need to consider the Food, Energy, and Water nexus. Developing IoT-based systems to address these concerns and the big data that stems from such systems are critical for the optimal provisioning and efficient utilization of the city's resources. In addition, providing smart solutions for transportation, healthcare, convenience, agriculture, and government are the main premises of a smart city. In this section we present smart city use cases that illustrate the use of semi-supervised learning to provide better services to the city's residents.
 
\subsection{Water}
\begin{figure}
	\begin{center}
		\includegraphics[width=0.5\textwidth]{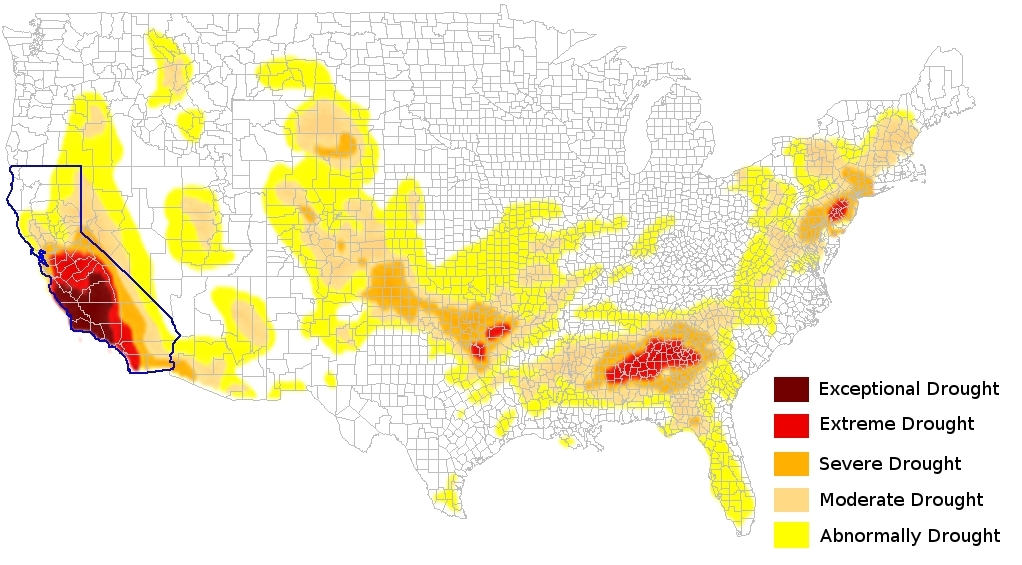}

	\end{center}
	\caption[The U.S. drought level]{The U.S. drought level in early 2017 and the situation of California. (Reproduced using R from  U.S. Drought Monitor data from Jan 3, 2017.)}
    \label{fig:drought}
\end{figure}

California has experienced an intense drought period in recent years. In early 2017, nearly all areas in California were under at least abnormal drought conditions. Also, in some areas the highest level of drought has been reported by the U.S. Drought Monitor\footnote{http://droughtmonitor.unl.edu} (See Figure \ref{fig:drought}). Analytics of big data from city temperature and humidity sensors, weather forecasts, prediction of water usage and the available water resources can help secure water for drought periods. Moreover, monitoring the level and quality of water in creeks using crowd-sensing data (e.g., the amount of trash, level of water, picture of trash, etc. in IBM CreekWatch) along with data from other IoT-based approaches such as smart water meters can help to achieve efficient and sustainable water provisioning. In this context, the images of trash in water can be used by a semi-supervised DRL system to identify the type of trash automatically and perform the required action at the location.

Using smart water meters can contribute to the fine-grained monitoring of water consumption at the house level as well as at the city level. Water consumption data at the house level can be analyzed by unsupervised clustering algorithms for abnormal and leakage detection. Imagine a scenario in which the household is on a trip for one week. An intelligent system based on DRL has learned that the water consumption between 5 and 6 p.m. on weekdays is in the range of $20\pm2$ liters when the household returns back home from work. The intelligent system also receives the location information about the household and determines that they are far from home at the time of the water meter reading which is on Monday at 5 p.m. At this time, a usage of 17 liters has been reported by the smart meter to the intelligent system. From the previous telemetry, the intelligent system can determine that the household is away from home and that the tap has not been turned off firmly. The intelligent system is trained so that the best rewarding policy is to stop the flow of water through the tap and notify the household accordingly.

This sort of intelligence has the potential of greatly impacting the whole city as what happened in the case of Kalgoorlie-Boulder, Australia where the installation of smart meters on the water pipelines led to early detection of leaks which in turn resulted in 12\% reduction in water consumption in one year \cite{boyle2013intelligent}.

\subsection{Energy} 
Energy conservation is a daily concern for people and energy utility service providers. Around one third of electricity usage is consumed by the residential sector in the European Union and the demand for energy is predicted to be double in the next decade. Energy providers nowadays can monitor consumers' energy usage profile and provide suitable feedback to decrease the high-peak power load using modern electricity meters (i.e., smart meters) that are installed on customers’ premises. 

Smart meters can also be connected to their smart home systems to cooperate with other devices toward energy management at the level of smart home using Appliance Load Monitoring (ALM). In this context, each electrical device is equipped with a smart power outlet. A semi-supervised DRL agent can observe its environment including the energy usage profile of electrical devices, the ambient temperature, the light intensity, and status of motion detectors to learn the best policies to turn off devices. The duration of the off period for the participating electrical devices can be considered as a reward function for the agent. However, this fine-grained level of ALM causes extra equipment cost and complexity. Instead, Non-Intrusive Load Monitoring (NILM) is an alternative approach that can extract the individual devices' usage from one aggregated electrical measurement at the scope of the whole house. This approach needs to be trained one time by the consumption data of individual appliances and their events (i.e., on or off) and time-stamps. A semi-supervised DRL agent can be utilized and integrated into this context aiming to keep the optimal power usage by controlling when to turn appliances on or off. Due to  the presence of many unlabeled data generated by NILM, the performance of the semi-supervised DRL agent is better than that of the supervised DRL agent.

The usage of smart energy in the context of smart grid has proven its payback as in the case of smart grid in Chattanooga, TN, using the smart grid helped in faster repairs after a severe storm outage in July 2012. This single incident helped in saving \$1 million \cite{talbot2017smart}.

\subsection{Agriculture} 

Agriculture activities are the main source for food production. Monitoring the soil parameters (e.g., moisture, minerals, etc.) powered by decision making processes, and consequently performing corrective actions by actuators (e.g., adding water or minerals), can lead to increased crops productivity.

Also, for producing healthy crops and efficiently growing plants, disease recognition and remedies is paramount. Plant disease recognition can be performed through disease recognition systems through various measurements. A viable approach is to identify diseased plants visually using a classification system based on images of the crops or their leaves. Farmers can install such systems on their smart devices to identify fruits and crops with anomalies. By combining those data with complementary data sources, the system can recommend remedies or pesticides to the farmers.

\section{Challenges and Future Directions}\label{sec:future_direction}

\subsection{Challenges}
Development of smart city applications supported by big data analytics is subject to several challenges that need to be addressed to achieve a reliable and accurate system. Some of the major challenges beyond the ones introduced by the 3V's include:
\begin{itemize}
\item Integrating big and fast data analytics: In a smart city context, there are many time-sensitive applications (e.g., smart vehicles) that need real-time or near real-time analytics of the stream of data. Such applications call for new analytic frameworks that support big data analytics in conjunction with fast data analytics. 

\item Preserving security and privacy: Data-driven machine learning approaches (e.g., deep learning) can be attacked by False Data Injection (FDI) which compromises the validity and trustworthiness of the system. Resilience against such attacks is a must for ML algorithms. Privacy preservation is another important factor since a large part of smart city data comes from individuals who may not prefer their data to be publicly available. ML algorithms should address these concerns to enable the wide acceptance of smart city systems by organizations and citizens.

\item On-device intelligence: Smart city applications also call for light-weight machine learning algorithms deployable on resource-constrained devices for hard real-time intelligence. This is also inline with the security and privacy preservation requirement since data is not transferred to the fog or cloud.

\item Big dataset shortage: Development and evaluation of smart city applications need real-world datasets which are not readily available for many application domains. It is required to confirm results based on simulated big data.

\item Context-awareness: Integrating contextual information with raw data is crucial to get more value of the data and perform faster and more accurate reasoning and actuation
\cite{perera2014context}. For example, detecting a sleepy face in a human pose detection system could lead to totally different actions in the contexts of driving a car and relaxing at home.

\end{itemize}

In addition, there are other challenges that affect the design of a smart city ecosystem such as integration of different analytic frameworks, distribution of analytic operations, and lack of comprehensive testbeds. 

\subsection{Future Directions}
The conventional analytic approach for IoT is to send raw data to the cloud for processing. However, this scheme is not effective and scalable for smart city deployments. Decentralization of data analytic computations is a new trend that aims to bring analytics closer to the fog and IoT devices \cite{tang2017incorporating}, \cite{han2015learning}. Here we list several promising future research directions in this regard.

\begin{itemize}

\item A trained model works well when the same feature set and distribution model forms the training and test data sets. By changing the distribution, the trained model needs to be rebuilt from new training data. For example, in a radio frequency based localization application (e.g., WiFi, BLE), the RSSI values for the same time and position in Android and iOS devices are different. The trained localization model on one platform can be transferred to the new platform without the need to collect RSSI values for other devices. Transfer Learning is a field of research that can help in such scenarios \cite{casale2015transfer}. 

\item Integration with semantic technologies is also a need for the development of smart city applications. The need stems from the interaction of those systems with citizens and the use of social media data. 

\item Intelligent virtual objects can be used in smart city services joint with DRL algorithms, considering each physical object has a virtual representation in the smart city and these VOs can learn, decide, and act autonomously.

\item Interacting with humans in a natural way is a critical need for the new generation of smart city systems since citizens are the main players in smart cities. The small size of mobile devices and wearables nowadays does not allow space for touch screens or keyboards. Instead, automatic speech recognition and natural language understanding is the most convenient way of interaction with these devices. 

\end{itemize}

\section{Conclusion}\label{sec:conclusion}

There are many machine learning algorithms that can be utilized to learn from the big data collected through a smart city's infrastructure. However, most traditional machine learning techniques assume a fixed training model and a static context. These assumptions do not apply to smart city applications where the environment and consequently the training data evolve over the time.

In this article, we addressed challenges and opportunities that arise when utilizing machine  learning to realize new smart city services. These challenges include: data-recycling, efficient sampling, and devising scalable models. We reviewed state-of-the-art methods and approaches that embrace smart city big data toward future cognitive smart cities. Then, a hierarchical framework was introduced to incorporate machine learning techniques in accord to the hierarchy of big data in smart city. We proposed a semi-supervised deep reinforcement learning framework to address the presented challenges and highlighted the position of the framework in various smart city application domains. Finally, we articulated several challenges and trending research directions for incorporating machine learning to realize new smart city services.

\bibliographystyle{IEEEtran}
\end{document}